\documentclass[11pt,twoside]{article}
\usepackage{asp2004}

\begin{document}
\title{The influence of binaries on galactic chemical evolution
}
\author{Erwin De Donder and Dany Vanbeveren}
\affil{Astrophysical Institute, Vrije Universiteit Brussel, Pleinlaan 2, 1050 Brussels, Belgium.
}

\begin{abstract}

\noindent Understanding the galaxy in which we live is one of the great intellectual challenges facing modern science. With the advent 
of high quality observational data, the chemical evolution modeling of our galaxy has been the subject of numerous studies in
the last years. However, all these  studies have one missing element which is \textsl{the evolution of close binaries}. Reason:
their evolution is very complex and single stars only perhaps can do  the job. (Un)Fortunately at present we know that the
majority of the observed stars are members of a binary or multiple system and that certain objects can only be  formed through
binary evolution. Therefore galactic studies that do not account for close binary evolution may be far from realistic. \\
\noindent Because of the large expertise developed through the years in stellar evolution in general and
binary evolution in particular at the Brussels Astrophysical Institute, we found  ourselves in a privileged
position to be the first to do chemical evolutionary simulations with the inclusion of detailed binary
evolution. The  complexity of close binary evolution has kept many astronomers from including binary stars
into their studies. However, it is not always the easiest way of living that gives you the most excitement
and satisfaction.\\

\noindent The paper is a 115 page review, published in New Astronomy Reviews 48, 861.
An offprint (with color figures) can be obtained via NARev or by sending a request to Dany
Vanbeveren, dvbevere@vub.ac.be.

\end{abstract}
\thispagestyle{plain}

\end{document}